\documentclass{ws-p8-50x6-00}

\begin{document}

\title{The QCD phase diagram for small chemical potentials}

\author{C.~Schmidt}

\address{Fakult\"at f\"ur Physik, Universit\"at Bielefeld, D-33615 Bielefeld,
  Germany\\
E-mail: schmidt@physik.uni-bielefeld.de}


\maketitle

\abstracts{
We compute derivatives of thermodynamic quantities with respect to $\mu$, at
$\mu=0$ for 2 and 3 flavors of degenerate quark masses. This allows us to
estimate the phase transition line in the $T,\mu$ plane and quantify the
influence of a non vanishing chemical potential on the equation of state by
computing lines of constant energy, pressure and density. Moreover we evaluate
the order of the QCD phase transition by measuring the Binder Cumulant of the
chiral condensate. This gives access to the chiral critical point on the
phase transition line.
}

\vspace*{-8cm}\hspace*{7.5cm} {\Large BI-TP 2002/26} \vspace*{7cm}

\section{Introduction}
To understand recent heavy-ion collisions, it is mandatory to study the QCD
phase diagram for high temperatures and small chemical potentials. It is well
known that the order of the phase transition is strongly dependent on quark
masses and chemical potential.
\begin{figure}[t]
\begin{center}
\epsfxsize=13pc
\epsfysize=8.5pc
\epsfbox{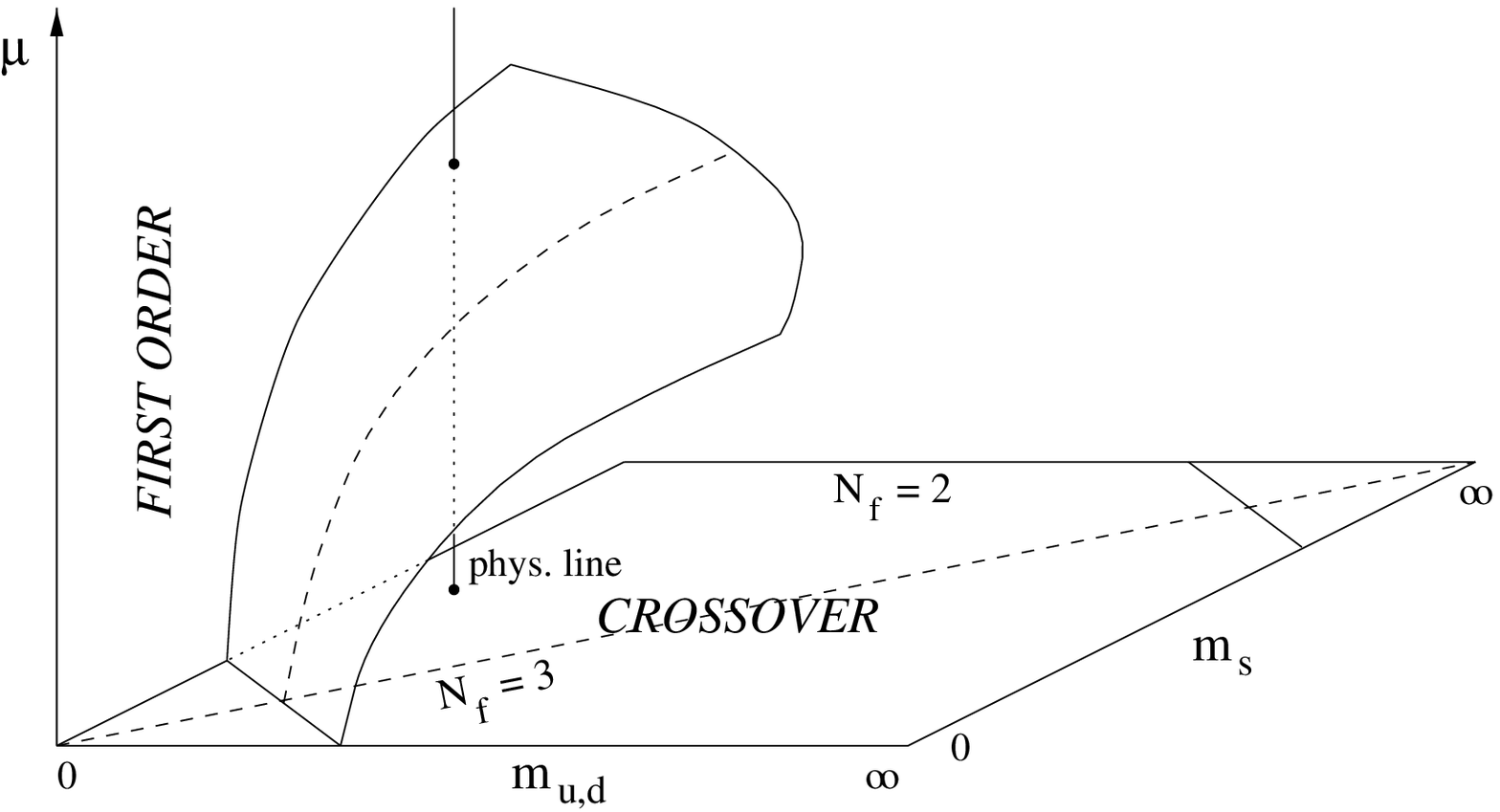}
\epsfxsize=13pc
\epsfbox{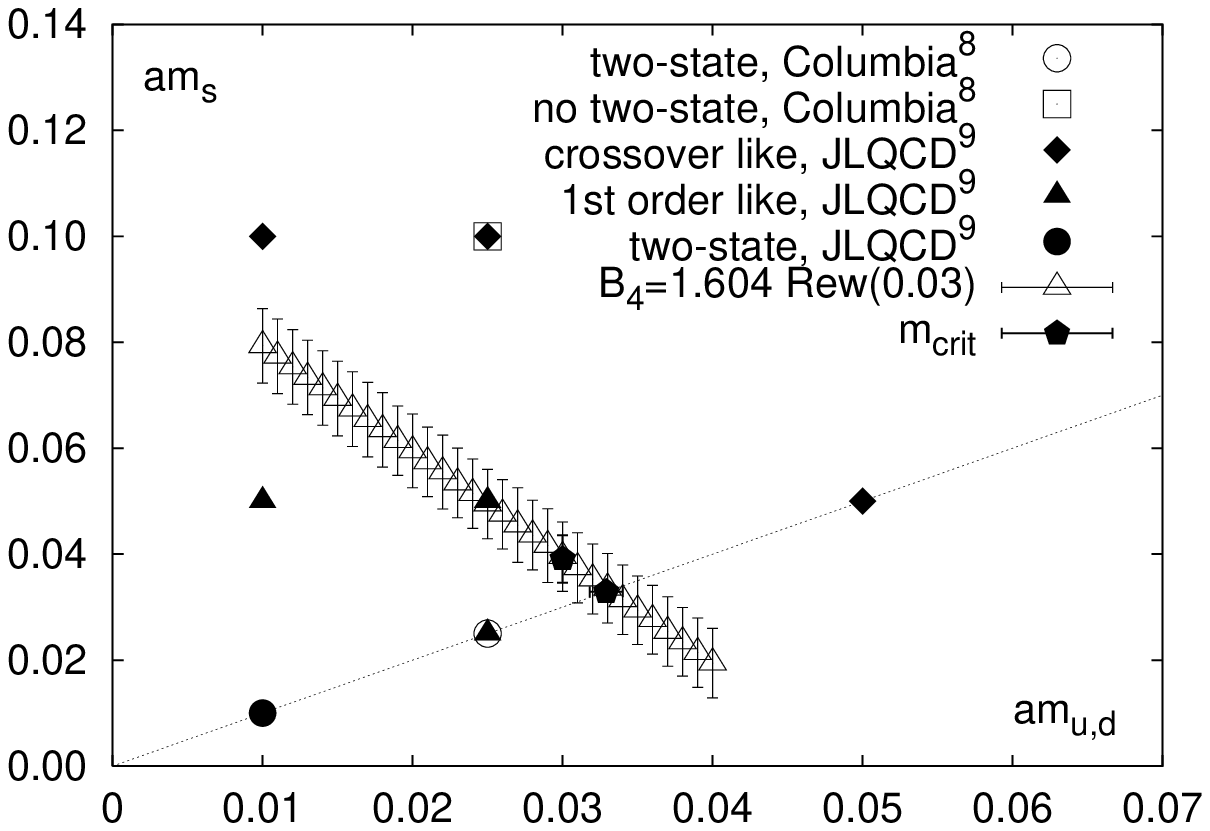}
\caption{Phase boundary for the first order regime of the thermal phase
  transition for non-vanishing $\mu$ (left), and the line of second order phase
  transitions for standard staggered fermions at $\mu=0$ (right).}
\end{center}
\end{figure}
In the space of two degenerate light quark masses $m_{u,d}$, one heavier mass
$m_s$ and the quark chemical potential $\mu$, one expects a critical surface,
which bends over the quark mass plane and separates the regime of first order
phase transitions from the crossover regime as shown in fig. 1.

To determine the order of the phase transition we compute the Binder Cumulant of
the chiral condensate, $B_4$, at $\mu=0$ for several quark masses at the
pseudo-critical coupling $\beta_c(m)$. We define $\beta_c$ as the peak position of
the chiral susceptibility. $B_4$ is given by
\begin{equation}
  B_4={\left<(\delta\bar\psi \psi)^4\right>_{\beta_c,m}
    \over
    \left<(\delta\bar\psi \psi)^2\right>^2_{\beta_c,m}},
    \qquad
    \delta\bar\psi\psi = \bar\psi\psi-\left< \bar\psi\psi
    \right>.
\end{equation}
This quantity is a renormalization group invariant quantity, with a universal
value at $m=m_{crit}$.  The universality class is that of the 3d Ising model with
$B_4=1.604$\cite{Gav94,schmidt,b4ising}. The critical surface is then given
by the surface of constant $B_4=1.604$.

To explore the regime of $\mu \ne 0$ we compute derivatives
of transition temperature, pressure, energy density, quark number density and
Binder Cumulant with respect to $\mu$. Direct Monte Carlo simulations for
$\mu\ne0$ are not possible due to the sign problem. We therefore use a
reweighting method at $\mu=0$ based on Taylor expansion of the fermion
determinant and all observables up to order $\mu^2$. To be more precise, the
expansion is given in terms of $\mu/T$ as the lattice chemical potential is
given in units of the lattice cut-off, $\mu^{latt}=a\mu^{phys}\propto \mu/T$.

\section{Reweighting in quark mass and chemical potential}
Ferrenberg and Swendsen's reweighting method is a very useful technique to
investigate critical phenomena. Multi parameter reweighting was first applied
to the problem of finite density QCD in ref. 4. We use the Taylor
expanded reweighting formula up to order $\mu^2$, which is given
by\cite{allton}
\begin{equation}
\left< O \right>_{(\beta,\mu)} = {
  \left< \left( O_0+O_1\mu+O_2\mu^2 \right) \exp\left\{
      R_1\mu+R_2\mu^2\right\} \exp\left\{-\Delta
      S_g\right\} \right> \over
\left< \exp\left\{R_1\mu+R_2\mu^2\right\}\exp\left\{-\Delta
      S_g\right\} \right>}.
\end{equation}
Here we have
\begin{equation}
R_i=\left.{N_f\over 4i!}{\partial^i \ln \det M (\mu) \over
    \partial \mu^i} \right|_{\mu=0}
\qquad \mbox{and} \qquad
O_i={1\over i!}\left.{\partial^i O (\mu) \over \partial \mu^i}
\right|_{\mu=0}
\end{equation}
We calculate the reweighting operators $R_i$ using stochastic estimators. One
can easily deduce the corresponding reweighting operators for mass
reweighting. We have control over the sign problem through the odd reweighting
operators $R_{2j+1}$, which are purely imaginary. The phase, $\Theta$, of the
fermion determinant, $\det M=|\det M|\exp\{i\Theta\}$, is in leading order
given by $\Theta=\mu\,\mbox{Im}R_1$. From this we find increasing phase
fluctuations for increasing volume, number of flavors, $n_f$, and for
decreasing $m$ and $T$.
\begin{figure}[t]
\begin{center}
\epsfxsize=13pc
\epsfysize=13pc
\epsfbox{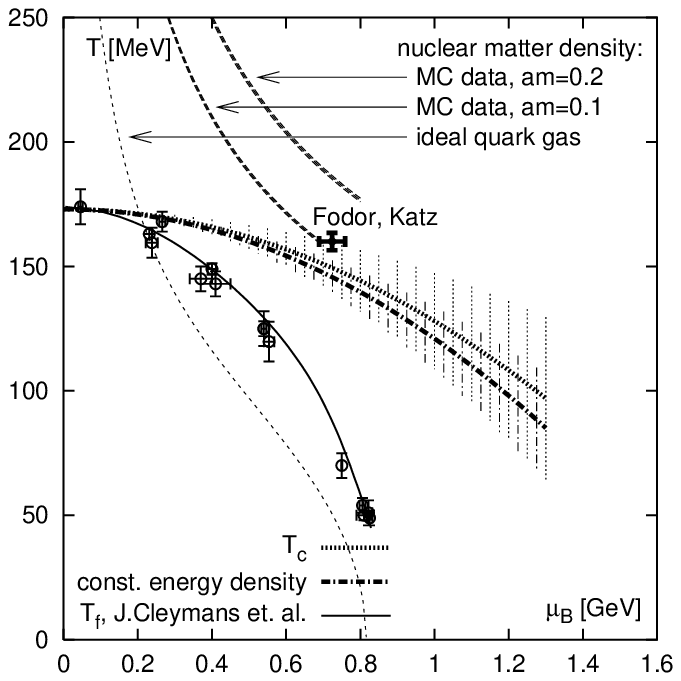}
\epsfxsize=13pc
\epsfysize=13pc
\epsfbox{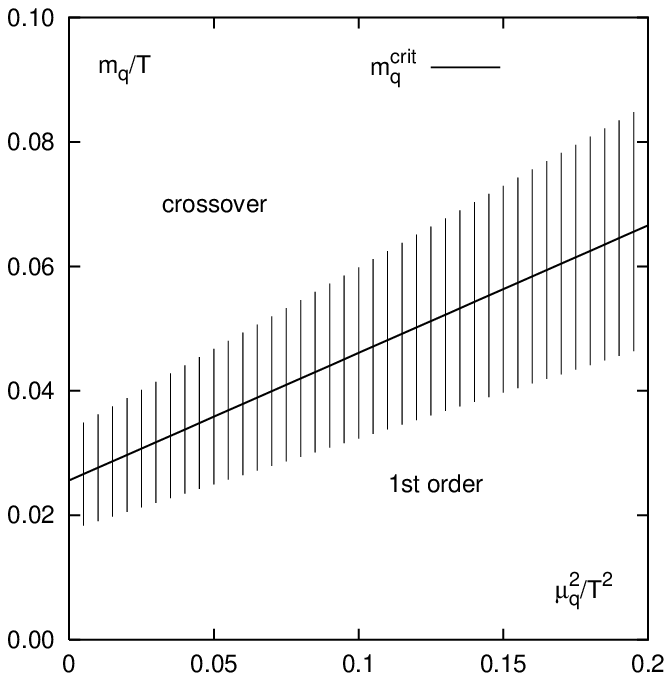}
\caption{The transition line in 2-flavor QCD together with lines of constant energy,
  nuclear matter density and the freeze-out temperature (left) and a sketch of the
  phase diagram in 3-flavor QCD (right).}
\end{center}
\end{figure}

\section{The transition to the QGP and the equation of state}
In 2-flavor QCD the transition from a hadron gas to the QGP will be a continuous
but rapid crossover for non vanishing quark masses and small values of $\mu$.
In order to estimate the transition line we performed simulations with 2
flavors of improved fermions\cite{peikert} on a $16^3\times4$ lattice and quark
masses of $am=0.1$ and $am=0.2$.
From the peak positions of the Polyakov loop susceptibility and the chiral
susceptibility we find $d^2 \beta_c/d \mu^2=-1.07(24)$ and $d^2 \beta_c/d
\mu^2=-1.10(68)$ respectively, which agrees with results of
ref. 7. To set the physical scale we use the string tension and
obtain $a(d\beta/da)=-2.08(43)$ at $(\beta,m)=(3.65,0.1)$. Combining these
results yields
\begin{equation}
T_c{d^2T_c\over d\mu_q^2} = -{1\over N_t^2}{d^2\beta_c \over
  d\mu^2} \left(a{d\beta \over a}\right)^{-1} =-0.14(7).
\end{equation}
Derivatives of the pressure $p$ and the interaction measure $\epsilon-3p$ are
related to the quark number density $n_q$ and quark number susceptibility
$\chi_s=\partial n_q/\partial\mu_q$ in the following way\cite{gottlieb}:
\begin{eqnarray}
{\partial (p/T^4) \over \partial \mu_q} &=& {1\over VT^3} {\partial \ln Z
  \over \partial \mu_q}={n_q \over T^4} \\
{\partial^2 (p/T^4) \over \partial \mu_q^2} &=& {1\over VT^3} {\partial^2 \ln
  Z \over \partial \mu_q^2} = {\chi_s \over T^4} \\
 {\partial^2 \over \partial \mu_q^2} \left( {\epsilon-3p \over T^4} \right)
  &\approx& {1\over T^4} {\partial \chi_s \over \partial \beta} \left( {1\over
  a}{\partial a \over \partial \beta} \right)^{-1}.
\end{eqnarray}
We obtain $T^2 \partial^2 (p/T^4) / \partial\mu_q^2\approx  0.69$ and  $T^2
\partial^2 (\epsilon/T^4) / \partial\mu_q^2  \approx 10.6$ at $\beta_c$ and
$am=0.1$. In the RHIC regime of $\mu_q/T_c \approx  0.1$ the deviation of
$p/T^4$ and $\epsilon/T^4$ from  results at $\mu=0$ thus is only a 1\%
effect. From the second derivatives of $p$ and $\epsilon$ with respect to $\mu$
together with the derivatives with respect to $T$, we calculate the lines of
constant pressure and energy density, which are within errors parallel to the
phase transition line:
\begin{equation}
T{dT\over d(\mu^2_q)}=\left\{\begin{array}{l}
                        -0.106(21):\mbox{(pressure)}\\
                        -0.081(20):\mbox{(energy)}\end{array}\right.,
\qquad T_c{dT_c\over d(\mu^2_q)}=-0.07(3).
\end{equation}
The quark number density is zero at $\mu=0$. For $\mu \ne 0$ It can be
estimated via $n_qa^3=\mu_qa\chi_Sa^2$. This is $n_q/T^3=0.693(5)\mu_q/T$ and
$n_q/T^3=0.490(4)\mu_q/T$ at $am=0.1$ and $am=0.2$, which translates into
roughly 9\% and 6\% of nuclear matter density at the RHIC point. All these
results are summarised in fig. 2 (left). In addition we also show the freeze-out
temperature\cite{cleymans}. This suggest that at sufficiently large $\mu_B$ there
is the chance to observe experimentally a strongly interacting hadron gas
phase, whereas for small $\mu_B$ hadrons seem to freeze out right after the phase
transition from QGP to the hadronic phase.

\section{The chiral critical point}
To determine the critical surface shown in fig. 1(left) we performed
calculations with much lighter quark masses. For 3 flavors of improved
fermions\cite{peikert}, a mass value of $am=0.005$ and a volume of $12^3\times
4$, we measured the reweighting operators $R_1, R_2$ needed for reweighting in
$m$ and $\mu$. At present we have five $\beta$-values, with a total number of
6100 trajectories. For $am=0.01$ and $V=16^3\times 4$ we use $R_1=\bar\psi
\psi$ for mass reweighting. To determine the critical mass value $\bar m$, we
compute $B_4$ as a function of $m$. The two different volumes have an
intersection point near the value of $B_4=1.604$, i.e. 3d Ising universality
class. Due to the large errors we give an upper bound for the critical mass
only, which is $a\bar m<0.0075$, or in terms of the pion mass $\bar m_{PS} <
190 MeV$.

For increasing quark chemical potential $\mu_q$ we find a decreasing Binder
Cumulant. From the two partial derivatives $\partial B_4/\partial (am)$ and
$\partial B_4/\partial (a^2\mu^2)$, which we get from straight line fits of
the reweighted data, and the assumption that $\partial B_4/\partial (am)$ is
constant in $am$, one can compute the quantity $\partial (a\bar m)/\partial
(a^2\mu^2)$. The first derivatives $\partial B_4/\partial (a\mu)$ and
$\partial (a\bar m)/\partial (a\mu)$ vanish because of symmetry reasons. A
jackknife analysis yields $a^{-1}\partial \bar m/\partial (\mu^2)=0.82(23)$,
or equivalently $T\partial \bar m/\partial (\mu^2)=0.21(6)$. We thus find that
the critical quark mass, $\bar m$, increases with increasing chemical potential
as anticipated in fig. 1 (left),
\begin{equation}
\bar m(\mu) \approx m_{3}  + 0.21\, \mu^2/T.
\end{equation}
Here $m_3$ is the critical quark mass for 3 degenerate flavor at $\mu=0$, which
turns out to be roughly twice as large as the physical light quark masses,
$m_{3}\approx 2\,m_{u,d}^{phys}$. This result is shown in fig. 2 (right).

Based on these results we can give a first estimate for the location of the
chiral critical point, i.e. the second order endpoint in the phase diagram
shown in fig. 2 (left). To do so we use $m_{u,d}^{phys}/T=0.016$, and assume
that the curvature of the critical surface is the same for 3 and 2+1-flavor. In
eq. 9 we should then replace $m_3$ by $m_{21}\approx4\,m_{u,d}^{phys}$, which can
be deduced from fig. 1 (right). We are interested in that point on the critical
surface that corresponds to the physical ratio of quark masses,
i.e. $R=m_s^{phys}/m_{u,d}^{phys}=24$. From eq. 9 we then find
$\mu_q/T_c\approx1.25$, which corresponds to $\mu_B\approx 3.75T_c \approx 650$
MeV and roughly coincides with the estimate given in ref. 12. Nevertheless our
present uncertainties do not yet allow a reasonable error estimate on this
value.

\section*{Acknowledgments}
The work has been supported by the DFG under grant FOR 339/1-2. I thank the
members of the Bielefeld-Swansea collaboration\cite{allton} for giving me the
opportunity to present our new preliminary results prior to publication.

\end{document}